\documentclass[twoside]{article}
\pdfoutput=1
\usepackage[accepted]{aistats_preprint}

\usepackage[round]{natbib}

\usepackage{etoolbox}
\makeatletter
\patchcmd\@combinedblfloats{\box\@outputbox}{\unvbox\@outputbox}{}{\errmessage{\noexpand patch failed}}
\makeatother

\long\def\comment#1{}
\makeatletter
\long\def\barenote#1{
    \insert\footins{\footnotesize
    \interlinepenalty\interfootnotelinepenalty 
    \splittopskip\footnotesep
    \splitmaxdepth \dp\strutbox \floatingpenalty \@MM
    \hsize\columnwidth \@parboxrestore
    {\rule{\z@}{\footnotesep}\ignorespaces
      #1\strut}}}
\makeatother

\makeatletter
\long\def\shrinkbox{\@ifnextchar[{\@mgnshrinkbox}{\@shrinkbox}}

\def\@shrinkbox#1{
	\begin{minipage}{\columnwidth}\begin{tabbing}
	#1
	\end{tabbing}\end{minipage}}

\def\@mgnshrinkbox[#1]#2{
   \setbox\@tempboxa\hbox{\unskip\begin{minipage}{\columnwidth}\begin{tabbing}
		#2
		\end{tabbing}\end{minipage}\unskip}
   \dimen0=\wd\@tempboxa
   \advance\dimen0 by #1
   \dimen1=\ht\@tempboxa
   \advance\dimen1 by #1
   \advance\dimen1 by #1
   \makebox[\dimen0]{\vbox to \dimen1{\vfil \box\@tempboxa \vfil}}}
\makeatother

\renewcommand{\eqref}[1]{eq.~\ref{eq:#1}}

\DeclareMathOperator*{\argmax}{arg\,max}

\renewcommand{\eqref}[1]{eq.~\ref{eq:#1}}

\usepackage{overpic}
\usepackage{bm}

\newcommand{\be}{\begin{equation}}
\newcommand{\ee}{\end{equation}}

\usepackage{booktabs}
\usepackage{amsfonts,amsmath,amssymb,amsthm} 
\usepackage[colorlinks=true,allcolors=black]{hyperref}

\begin{document}

%

%

\twocolumn[

\aistatstitle{Unifying and generalizing models of neural dynamics during  decision-making}

\aistatsauthor{ David M. Zoltowski \And Jonathan W. Pillow \And  Scott W. Linderman }

\aistatsaddress{ Princeton University \And  Princeton University \And Stanford University } ]

\begin{abstract}
An open question in systems and computational neuroscience is how neural circuits accumulate evidence towards a decision. Fitting models of decision-making theory to neural activity helps answer this question, but current approaches limit the number of these models that we can fit to neural data. Here we propose a unifying framework for modeling neural activity during decision-making tasks. The framework includes the canonical drift-diffusion model and enables extensions such as multi-dimensional accumulators, variable and collapsing boundaries, and discrete jumps. Our framework is based on constraining the parameters of recurrent state-space models, for which we introduce a scalable variational Laplace-EM inference algorithm. We applied the modeling approach to spiking responses recorded from monkey parietal cortex during two decision-making tasks. We found that a two-dimensional accumulator better captured the trial-averaged responses of a set of parietal neurons than a single accumulator model. Next, we identified a variable lower boundary in the responses of an LIP neuron during a random dot motion task. 
\end{abstract}

\section{Introduction}

Evidence accumulation is central to many models of perceptual decision-making \citep{gold2007neural,ratcliff2008diffusion,ratcliff2016diffusion}. However, despite progress in identifying neural correlates of decision-making
it remains an open question how evidence accumulation is implemented in the brain \citep{brody2016neural}.  One approach to address this question is to formulate models of decision-making behavior as generative models of neural activity \citep{ditterich2006stochastic,bollimunta2012neural, latimer2015single, zoltowski2019discrete}. Fitting these models to single-trial neural responses during decision-making tasks provides a direct test of how well the theorized model explains neural dynamics.  

However, there are a number of decision-making models and features that are challenging to fit to neural activity using existing approaches. Most models of decision-making behavior rely on either 1) approximations of analytic solutions of the joint distribution of a binary choice and decision time \citep[e.g.]{wiecki2013hddm}; or 2) numerical solutions to stochastic differential equations with boundary constraints \citep[e.g.]{brunton2013rats}.  Analytic solutions are not available for models of neural activity where there are observations throughout the decision-period, and numerical solutions are limited to low-dimensional accumulator models. Thus, there is a clear need for tractable methods for neural decision-making models that can accommodate multiple choice options and/or input streams \citep{churchland2008decision,brunton2013rats}, and choice dynamics governed by multi-dimensional accumulators \citep{scott2015sources}.  

Here we propose a general framework for fitting decision-making models to neural activity.\footnote{A Python package for fitting the decision-making models in this paper to neural data is available at \url{https://github.com/davidzoltowski/ssmdm}.} Our key observation is that many models of perceptual decision-making can be formulated as recurrent switching linear dynamical systems (rSLDS)  models \citep{linderman2017bayesian} with appropriate constraints. This allows us to instantiate and fit a number of models of interest including the classic drift-diffusion model \citep{ratcliff2008diffusion}, one- and multi-dimensional accumulator models \citep{gold2007neural,brunton2013rats}, and the ramping and stepping models introduced in \cite{latimer2015single}. The framework naturally includes collapsing boundaries and trial history effects \citep{oconnell2018bridging,urai2019choice}. Importantly, it also enables new extensions to models with variable, probabilistic boundaries and with non-constant boundary dynamics \citep{resulaj2009changes,evans2018synaptic}, which are not naturally described by traditional methods that assume constant, absorbing boundaries. 

\begin{figure*}[t]
    \centering
    \includegraphics[width=\textwidth]{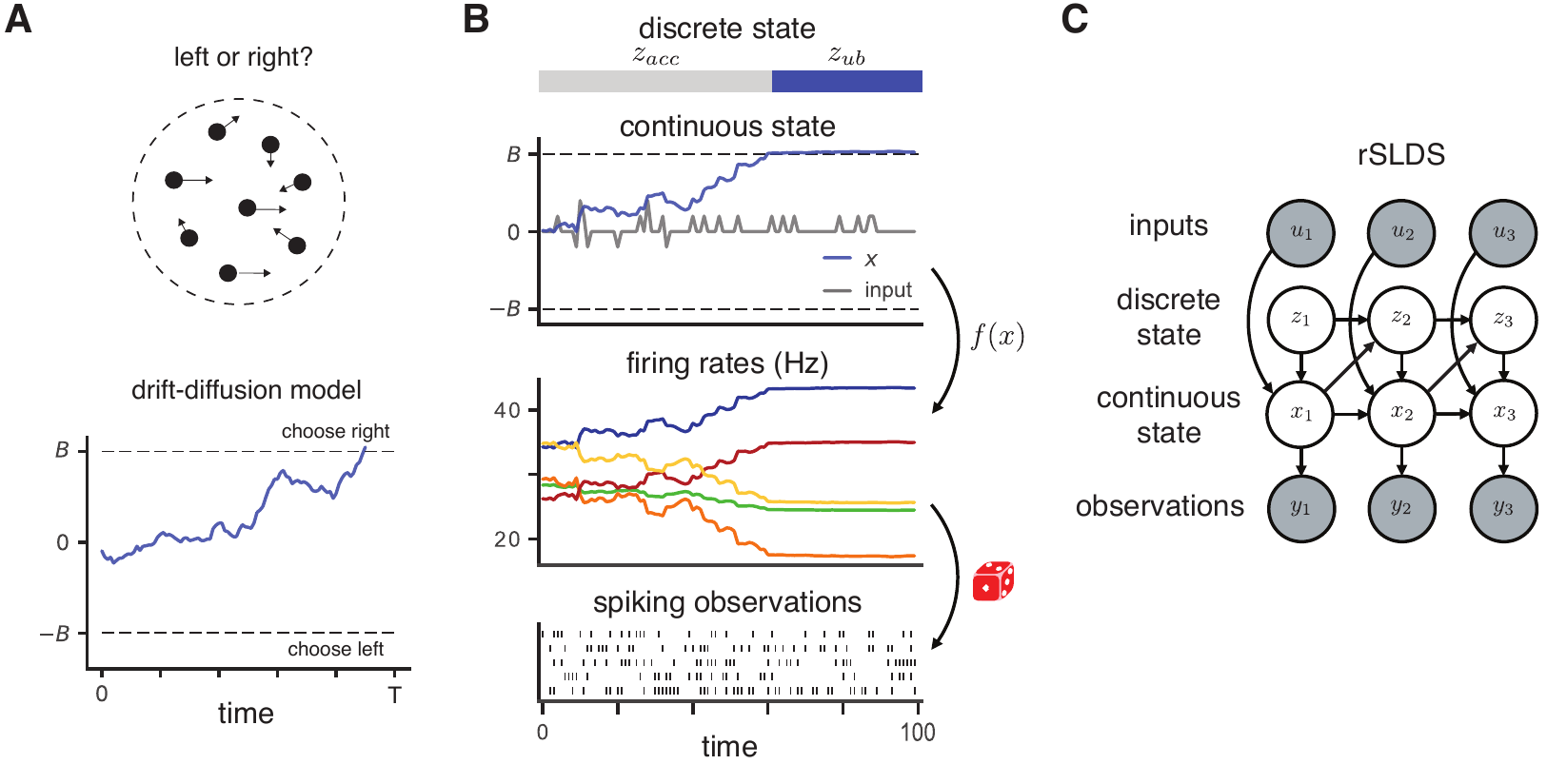}
    \caption{Constraining rSLDS to follow accumulation-to-bound dynamics. \textbf{A.} An example perceptual decision-making task is determining the direction of a cloud of randomly moving dots. The expected value of the motion direction is either to the left or right. The drift-diffusion model (DDM) is commonly used to explain decisions in these types of tasks. \textbf{B.} An rSLDS with constraints on the parameters can implement accumulation-to-bound. A simulated trial from this model in is shown. The continuous latent variable accumulates input in the accumulation state before switching to the upper boundary state after it crosses the upper threshold. Firing rates are generated from an affine mapping of the continuous variable and Poisson spike counts are generated. \textbf{C.} The rSLDS is a switching linear dynamical system with additionaly dependencies between~$x_t$ and~$z_{t+1}$, as shown in this graphical model.}
    \label{fig:fig1}
\end{figure*}

We also develop a variational Laplace-EM algorithm for inference in rSLDS models. This method combines variational and Laplace approximations over the discrete and continuous latent variables, respectively. We assess the algorithm's ability to recover latent states and model parameters in synthetic data, and show that we can fit new decision-making models to Poisson spike train data. We find that multi-dimensional accumulator models outperform existing methods for fitting neural responses during decision-making, suggesting new paths forward for the theory of neural decision-making. 

\section{Background}

We briefly review decision-making models for behavioral and neural data, and the recurrent switching linear dynamical systems we will use to represent, fit, and generalize them.

\subsection{Perceptual decision-making}

Perceptual decision-making is the process of using sensory information to make a categorical choice \citep{hanks2017perceptual}. A classic example is determining the net motion direction of a cloud of randomly moving dots (Figure~\ref{fig:fig1}A). 
The accumulation of evidence over time in favor of a decision until a threshold or boundary is crossed (accumulation-to-bound) is a key process in models of such perceptual decisions  \citep{gold2007neural,ratcliff2008diffusion}. 

A simple example model of perceptual decisions is the drift-diffusion model (DDM). The DDM is a 1-dimensional dynamics model of a binary choice (Figure~\ref{fig:fig1}A). It states that a scalar variable $x$ evolves in time according to a biased random walk. In discrete time this is formalized as
\begin{equation}
x_t = x_{t-1} + \beta_c + \epsilon_t, \quad \epsilon_t \sim \mathcal{N}(0, \sigma^2).
\end{equation}
where $\beta_c$, the drift term, corresponds to the strength of sensory evidence and noise $\epsilon_t$ corresponds to noise in the sensory input. Once $x$ crosses an upper or lower threshold (boundary) the decision-process stops and the variable $x$ is fixed to one of the two boundary values. The choice produced by the model depends on which of the upper or lower thresholds is hit; each boundary represents one of the two choices. Depending on the parameterization, the initial value may be set to a constant $x_0$ or drawn from an initial distribution $x_0 \sim \mathcal{N}(\mu_0, \sigma^2_0)$. Many variants of drift-diffusion models exist; see \cite{ratcliff2016diffusion} for a recent review. 

\subsection{Relating neural activity to decisions}

Neural correlates of decision-making have been identified in numerous brain regions \citep{gold2007neural,brody2016neural,hanks2017perceptual}. While the neural correlates of decisions have traditionally been observed when averaging neural activity across many decisions, a prominent line of work has attempted to relate single-trial neural responses to decision-making dynamics \citep{ditterich2006stochastic,churchland2011variance,bollimunta2012neural,latimer2015single,hanks2015distinct,zoltowski2019discrete}. An example of this is the ramping model introduced in \cite{latimer2015single} that posits that a modified DDM without a lower boundary underlies the firing rates of individual neurons in the monkey parietal cortex, following theoretical models \citep{mazurek2003role}. 

\subsection{Recurrent switching state-space models}

An rSLDS is an extension of a linear dynamical system model that approximates nonlinear dynamics with a discrete set of linear regimes \citep{linderman2017bayesian,nassar2019treestructured}. The generative process is as follows. At each time point, the dynamics are given by 
\be
x_{t} = A_{z_t} x_{t-1} + V_{z_t} u_t + b_{z_t} + \epsilon_t, \quad \epsilon_t \sim \mathcal{N}(0, Q_{z_t})
\ee
where $x_t$ is the continuous state, $z_t$ is one of $K$ discrete states, and $u_t$ is the input at time $t$. There are separate dynamics parameters $A_k, V_k, b_k$ for each discrete state $k \in \{1, 2, \ldots, K\}$. The key feature that makes the models ``recurrent'' is that transitions between discrete states depend on both the previous continuous and discrete latent variables. We parameterize the transition probabilities using a multi-class logistic regression 
\be
p(z_{t} \mid z_{t-1}, x_{t-1}) \propto \exp \left \{ \gamma (R_{z_{t-1}} + r \, x_{t-1}) \right \}
\label{eq:transitions}
\ee
with parameters $R_{k}$ for each discrete state and a vector $r \in \mathbb{R}^d$ describing the dependence on the previous continuous latent state. The hyperparameter $\gamma$ sets the sharpness of the decision boundaries. As $\gamma \to 0$, the transitions become uniform, and as $\gamma \to \infty$ they become deterministic.

In our applications, the most common observation distribution for our data is Poisson
\be
y_t \sim \operatorname{Poisson}( f(C x_t +  d ) \, \Delta_t)
\label{eqn:poisson}
\ee
where $f(x) = \log(1+\exp(x))$ is the softplus function and $\Delta t$ is the time bin size. However, it is possible to use other observation distributions for alternative types of data. The nonlinearity could also be changed to model accumulator tuning curves \citep{hanks2015distinct,depasquale2019accumulated} or saturating or accelerating relationships between the diffusion process and firing rates \citep{howard2018evidence,zoltowski2019discrete}. The graphical model of an rSLDS is shown in Figure~\ref{fig:fig1}C.

\section{Decision-making dynamics as constrained rSLDS}
Our key observation is that the dynamics of accumulation-to-bound models can be instantiated as constrained recurrent switching linear dynamical systems. We first illustrate this by writing a one-dimensional accumulation-to-bound as a constrained rSLDS. We then describe how this approach generalizes to multi-dimensional accumulators, ramping and stepping models, variable boundaries, and other features. 

\begin{figure*}
    \centering
        \includegraphics[width=1.0\textwidth]{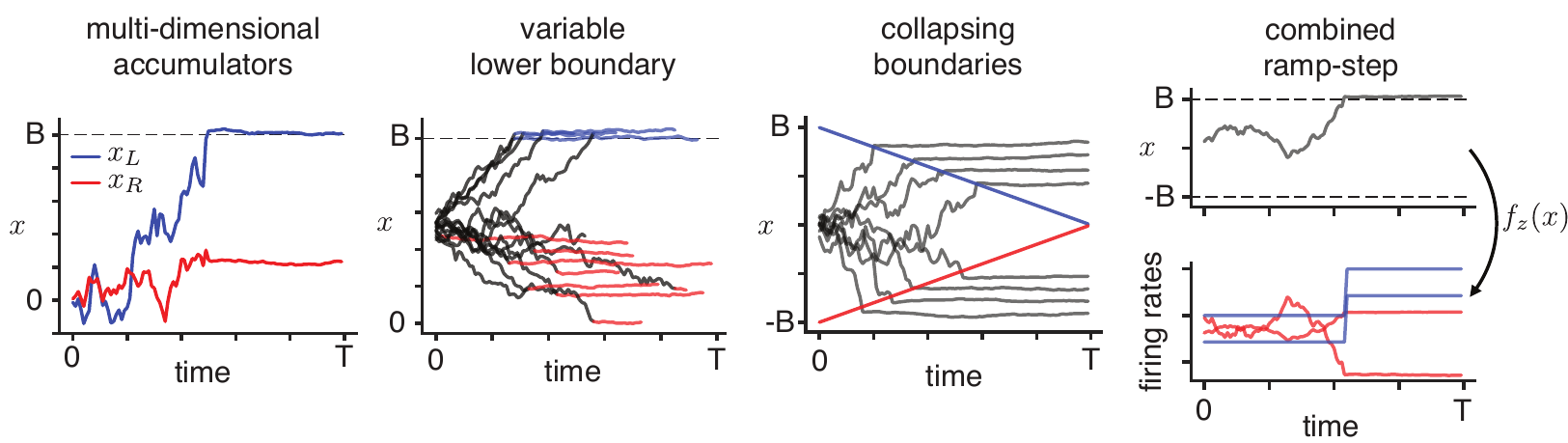}
    \caption{Examples of models in this framework. Constrained rSLDS models include multi-dimensional accumulators such as a 2D race accumulator model, accumulators with variable boundaries and collapsing boundaries, and a model with both discrete steps and ramps as special cases.}
    \label{fig:fig2}
\end{figure*}
\subsection{Illustrative example: one-dimensional accumulator model}
Consider a one-dimensional accumulation-to-bound model with upper and lower decision boundaries. This model has a continuous latent variable $x_t \in \mathbb{R}^1$ that starts near zero and accumulates sensory input until it reaches one of two decision boundaries at $\pm B$. The model has $K=3$ discrete states: an accumulation state $(z_t = \mathsf{acc})$, an upper boundary state $(z_t = \mathsf{ub})$, and a lower boundary state $(z_t = \mathsf{lb})$. 

In the accumulation state, the sensory input $u_t$ at each time point is accumulated via the following dynamics
\be
x_t = x_{t-1} + V_{{\mathsf{acc}}} u_t + \epsilon_t, \quad \epsilon_t \sim \mathcal{N}(0, \sigma_{{\mathsf{acc}}}^2)
\ee
where $V_{{\mathsf{acc}}}$ is a weight on the sensory input. In the boundary states there is no dependence on the input ($V_{\mathsf{ub}} = V_{\mathsf{lb}} = 0$) such that
\be
x_t = x_{t-1} + \epsilon_t, \quad \epsilon_t \sim \mathcal{N}(0, \sigma_{{\mathsf{ub/lb}}}^2)
\ee
and the diffusion variances $\sigma_{{\mathsf{ub}}}^2$ and $\sigma_{{\mathsf{lb}}}^2$ are set to small values relative to the scale of the dynamics. In all three states, $A = 1$ and $b = 0$. 

To implement the decision boundaries, we parameterize the transitions such that the discrete state switches from the accumulation to the boundary states once $x_{t-1}$ has crossed one of the boundaries. The transition parameters in equation~\ref{eq:transitions} depend on a boundary location $B$ parameter such that 
\begin{align*}
R_{\mathsf{acc}} = 
\begin{bmatrix}
~~0 \\ - B \\ - B
\end{bmatrix}\!\!, \, 
R_{\mathsf{ub}} = 
\begin{bmatrix}
-\infty  \\ ~~0 \\ -\infty 
\end{bmatrix}\!\!, \,
R_{\mathsf{lb}} = 
\begin{bmatrix}
-\infty \\ -\infty \\ ~~0
\end{bmatrix}\!\!,
\,
r = 
\begin{bmatrix} ~~0 \\ +1 \\ -1 \end{bmatrix} \!\!.
\end{align*}
Here we set $B = 1$ and $\gamma = 500$ to enforce sharp transition boundaries at~$\pm 1$. Figure~\ref{fig:fig1}B shows a simulation from this model with Poisson spike count observations, as described in equation~\ref{eqn:poisson}. 

\subsubsection{Soft decision boundaries}
We can implement soft decision boundaries where switches from the boundary state back to the accumulation state are allowed if $x_t$ falls below the boundary. Here the transition probabilities depend only on the previous value of the continuous latent variable with 
\begin{equation}
R_{\mathsf{acc}} = R_{\mathsf{ub}} = R_{\mathsf{lb}} = \begin{bmatrix}
~~0 \\ -B \\ -B
\end{bmatrix}\!\!, \; 
r = \begin{bmatrix} ~~0 \\ +1 \\ -1 \end{bmatrix}\!\!.
\end{equation}

\subsection{Multi-dimensional accumulator models}

Models with multiple accumulating dimensions are natural for settings with multiple input streams and choice options. Additional dimensions may also be useful for incorporating other task features such as context \citep{shvartsman2015theory}.
In this framework, it is straightforward to generalize the one-dimensional accumulator model to multi-dimensional accumulators by adding dimensions to the continuous and discrete states. For a $D$-dimensional accumulator, the continuous latent variable $x \in \mathbb{R}^D$ and the input $u \in \mathbb{R}^D$ are vectors. We set the dynamics, input, and covariance matrices $\{A_{\mathsf{acc}}, V_{\mathsf{acc}}, Q_{\mathsf{acc}} \}$ to be diagonal so that each dimension independently accumulates one stream of input. However, we can relax this assumption to have interactions in the latent space (non-diagonal $A_{\mathsf{acc}}$) or correlated noise (non-diagonal $Q_{\mathsf{acc}}$).

We set the discrete state transitions to follow ``race'' accumulator dynamics, where the different accumulator dimensions race to reach an upper boundary. The discrete state switches when one of the dimensions of $x$ crosses the boundary. As in the 1D model, the boundary states have zero dependence on the input and small dynamics variance. In this setup, there is one accumulation state and $D$ boundary states such that $K = D+1$. A set of simulated latent trajectories from this model is shown in Figure~\ref{fig:fig2}.

\renewcommand{\arraystretch}{1.25}
\begin{table*}[t!]
\centering
\caption{Decision-making models and features that fit in the framework.} 
\begin{tabular}{l l l} 
\\
\toprule
feature & component & details \\ 
\midrule 
leaky or unstable dynamics & continuous dynamics & learn $A_{\mathsf{acc}}$ \\
input-dependent noise & continuous dynamics &  $\sigma_\mathsf{acc}^2 = \sigma_a^2 + u_t^2 \sigma_s^2$ \\ 
history bias: drift & continuous dynamics & augment input $u_t = [u_t, c_{\textsf{prev}}]$ \\
relaxed boundary dynamics & continuous dynamics & learn $A_{\mathsf{lb}}, A_{\mathsf{ub}}$ and/or $\sigma^2_{\mathsf{lb}}, \sigma^2_{\mathsf{ub}}$  \\
variable drift & continuous dynamics & hierarchical model of $V_{\mathsf{acc}}$ across trials \\
variable boundaries & discrete transitions & decrease or learn scale parameter $\gamma$ \\
collapsing boundaries & discrete transitions & switch to $\mathsf{ub}$ or $\mathsf{lb}$ when $|x_{t-1}| > B - f(t)$ \\
history bias: start & initial state & $x_0 \sim \mathcal{N}(\mu_{z_0} + V_{c} \, c_{\textsf{prev}}, \sigma^2_{z_0})$ \\
multi-dimensional & all & increase dimensions \\
non-decision time & all & start in additional state $z_0 = \mathsf{nd}$ before $\mathsf{acc}$ \\
\bottomrule
\end{tabular}
\label{table:tab1}
\end{table*}
\renewcommand{\arraystretch}{1}

\subsection{Ramping and stepping models} 
The ramping model is a one-dimensional accumulator model with an upper boundary \citep{latimer2015single}. It has a constant drift for each stimulus category such that $u_t$ is a one-hot vector denoting the stimulus on each trial and $V_{\mathsf{acc}}$ is a vector with a drift for each category. The model has no lower boundary and the initial continuous state has mean $x_0 \in (-\infty, 1)$. This model can be written as a constrained recurrent state space model with $K=2$ discrete states. Alternatively, the stepping model from \citet{latimer2015single} posits that single-trial firing rates start in an initial state and may either step up or down at some point during the trial.

The recurrent state-space framework admits a generalization of the ramping and stepping models that can be either a ramp or a step depending on free parameters. In this model, the underlying latent follows 1D accumulation-to-bound dynamics with upper and lower boundaries and the emission mean parameter depends on the discrete state
\begin{equation}
    y_t \sim \operatorname{Poisson}(f(C x_t + d_{z_t}) \, \Delta t).
\end{equation}
The model is a ramp if the mean $d_{z_t}^{(n)}$ for a neuron $n$ is the same for each discrete state and $C^{(n)}$ is non-zero. If $C^{(n)} = 0$ and $d_{z_t}^{(n)}$ is different for each state, then the model steps to different firing rates when the discrete state switches. An illustration of an underlying ramp generating firing rates for two ramping and two stepping neurons is shown in Figure~\ref{fig:fig2}.  

\subsection{Variable and learned boundaries}

We may desire to relax the assumption of sharp boundaries governing the discrete transitions. As the scale parameter $\gamma$ is increased, the distribution over transitions flattens and the transitions may occur at values below or above the boundary (Figure~\ref{fig:fig2}). It is also possible to learn the parameters governing the transitions. In our experiments, we present a model with a sharp upper boundary and a learned lower boundary. 

\subsection{Other features} The modeling framework includes numerous other features in decision-making models, such as sensory-dependent vs. accumulation noise \citep{brunton2013rats}, collapsing boundaries and non-decision time \citep{ratcliff2016diffusion}, and trial-history effects that for example depend on the previous choice $c_{prev}$ \citep{urai2019choice}. We list them in Table~\ref{table:tab1} and describe input-dependent noise, collapsing boundaries, and relaxed boundary dynamics below. While we do not fit models with all of the following features in this paper, we present them here for completeness and we fit models with collapsing boundaries and trial-history effects to simulated data in Appendix~\ref{sec:appendix_experiments}. 

\textbf{Input-dependent dynamics variance~~}
In decisions with non-constant input, dynamics variability can be separated into input-driven and dynamics components \cite{brunton2013rats}. This corresponds to adding a term to the accumulation dynamics variance that is a function of the current input
\begin{equation}
    x_t = A_\mathsf{acc} x_{t-1} + V_\mathsf{acc} u_t + \epsilon_t, \quad \epsilon_t \sim \mathcal{N}(0, \sigma_a^2 + u_t^2 \sigma_s^2).
\end{equation}

\textbf{Collapsing boundaries~~}
Collapsing decision boundaries are a feature of decision-making models that can account for decision accuracy as a function of decision-time \citep{ratcliff2016diffusion,oconnell2018bridging}. Linearly collapsing boundaries can be implemented by adding a term to the transition probabilities that depends on time $t$ such that
\be
p(z_{t} \mid z_{t-1}, x_{t-1}) \propto \exp \left \{ \gamma (R_{z_{t-1}} + r \, x_{t-1} + W t )\right \}
\ee
where $W = [0, \beta, \beta]^\top$ (Figure~\ref{fig:fig2}). The parameter $\beta$ determines the slope of the linear boundary decrease. We note that the boundary dependence on time could also be a nonlinear function of $t$. 
See Appendix~\ref{sec:appendix_experiments} for more details and a simulated fit of the model.

\textbf{Relaxed boundary dynamics~~}
The models introduced above make strong assumptions that the dynamics in the boundary state are constant with nearly zero noise. However, we can naturally relax these assumptions by learning the dynamics matrices and variances in the boundary states. Relaxed boundary dynamics could correspond to working memory and have been used to describe changes of mind \citep{resulaj2009changes} and mouse escape behavior \citep{evans2018synaptic}. 

\subsection{Differences with decision-making models}

Our recurrent state-space formulation of accumulation to a boundary departs from some formulations of decision-making models in two ways. First, the continuous state $x_t$ can cross the boundary and fluctuate, rather than reaching a constant level after hitting the boundary. We consider this a feature because it enables modeling neural dynamics after the boundary is reached. However, using a small boundary variance and identity dynamics matrix mitigates this difference, as described above. Additionally, to enforce the continuous state to remain at the boundary we can threshold $x_t$ when passing it to the firing rate. 
The second difference is that the transitions are probabilistic. We also consider this a feature because it allows for generalizations such as variable boundary locations. However, we can set the transitions to be effectively deterministic using sharp transition boundaries, as described previously. 

\begin{figure*}[t]
    \centering
    \includegraphics[width=1.0\textwidth]{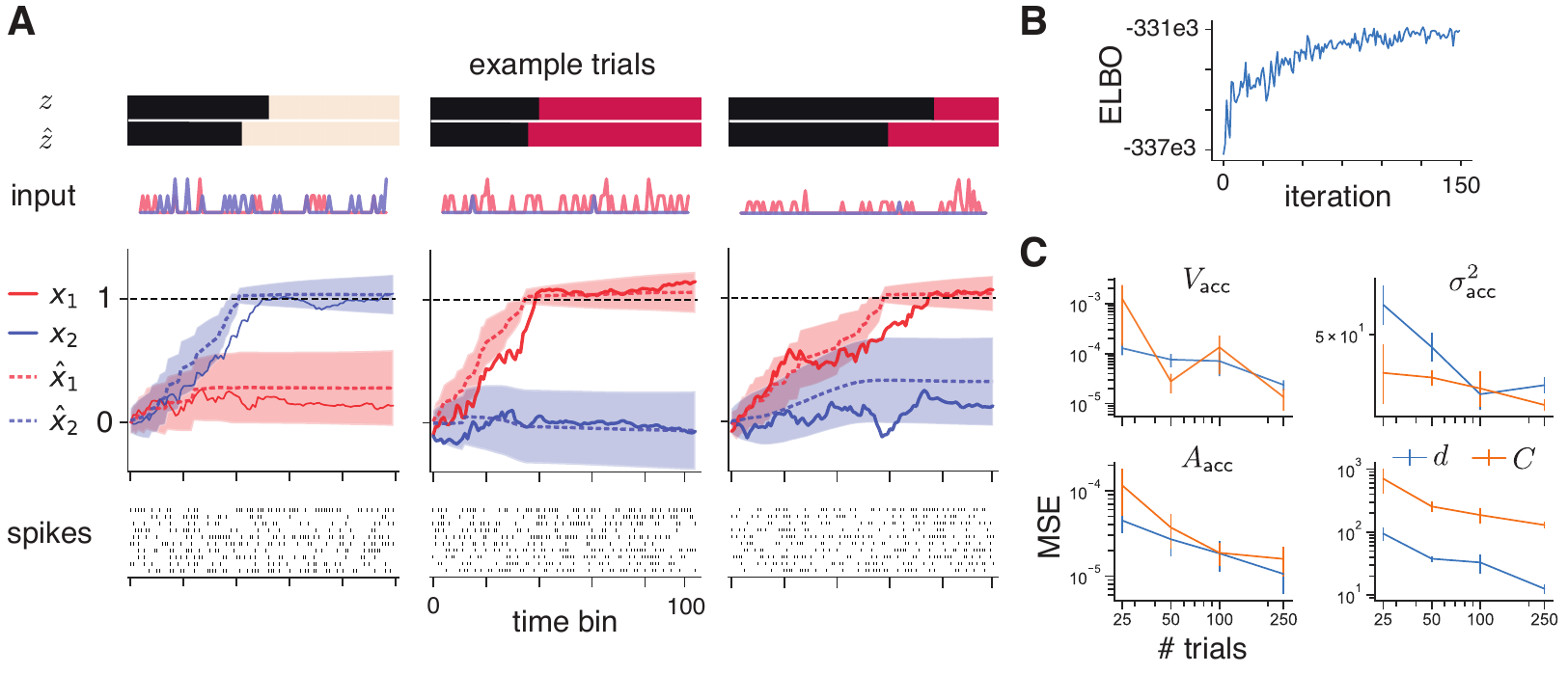}
    \caption{Simulated 2D race accumulator model with Poisson observations. \textbf{A.} The true ($z, x$) and inferred ($\hat{z}$, $\hat{x}$) discrete and continuous latent variables are similar. The colored bars (\textit{top}) indicate the true and inferred most likely discrete states at each time point. Shaded areas indicate two standard deviations of the posterior $q(x)$. \textbf{B.} The ELBO as a function of inference iteration. \textbf{C.} Mean MSE $\pm$ standard error between the true and inferred parameters across 10 simulated experiments for each number of trials. The different colored lines on the $V_\mathsf{acc}$, $\sigma^2_\mathsf{acc}$, and $A_\mathsf{acc}$ panels denote the parameters for the two different dimensions.}
    \label{fig:fig3}
\end{figure*}

\section{Inference}

Inference in rSLDS models via direct maximum likelihood scales exponentially in the length of the time-series and is therefore computationally intractable \citep{barberBRML2012}. A number of alternative methods have been developed for inference in rSLDS models including expectation-propagation~\citep{barber2006expectation} and augmented Gibbs sampling algorithms~\citep{linderman2017bayesian,nassar2019treestructured}. However, our use of multinomial logistic regression for the transition distributions and Poisson observations render the model non-conjugate. Further, we found that inference using black-box variational inference (BBVI) approaches \citep{archer2015black, gao2016linear, linderman2019hierarchical} was difficult given the strong boundary parameter constraints (see Appendix~\ref{sec:appendix_experiments}).   


We therefore developed an approximate inference algorithm that exploits information about the parameter constraints yet also has favorable scaling. The method combines a variational approximation of the posterior over the discrete states and a Laplace approximation of the posterior over the continuous states. Here we present an overview of the method, variational Laplace-EM (vLEM), with additional details in Appendix~\ref{sec:appendix_inference}.

We first introduce a factorized approximate posterior over the latent variables $p(z,x \,|\, \theta, y) \approx q(z) q(x)$, where $q(z)$ is a variational approximation and $q(x)$ is computed via a Laplace approximation. This admits a lower-bound on the marginal likelihood (ELBO)
\begin{multline}
 \log p(y \mid \theta)  \geq \mathcal{L}_q(\theta) \\ 
 =  \mathbb{E}_{q(z) q(x)}[\log p(z, x, y \mid \theta)] \\
 \quad - \mathbb{E}_{q(z)}[\log q(z)] - \mathbb{E}_{q(x)}[\log q(x)].
\end{multline}
We alternate between updating the two approximate posteriors and the model parameters in three steps. 

First, the discrete state approximate posterior is updated using the optimal coordinate ascent variational inference update \citep{bishop2006pattern,blei2017variational}
\begin{equation}
q^\star(z)  \propto \exp ( \mathbb{E}_{q(x)}[\log p(x, z, y | \theta)] ).
\end{equation}
This step locally maximizes the ELBO given a fixed $q(x)$ and model parameters $\theta$. We compute the expectation in the update using Monte Carlo samples from $q(x)$. Conditioned on these samples, the posterior $q^\star(z)$ has the same factor graph as an HMM and we can use the forward-backward algorithm to compute the posterior distributions over the discrete states and the marginal likelihood. 

The second step is to update $q(x)$ using a Laplace approximation around the most likely latent path $x^\star$ \citep{paninski2010new,macke2011empirical,macke2015estimating}. That is, we set $q^\star(x)$ to be 
\begin{equation}
q^\star(x)  = \mathcal{N}(x^\star, -H^{-1}) 
\end{equation}
where 
\begin{align}
x^\star &= \argmax_x \, \mathbb{E}_{q(z)}[\log p(x, z, y \mid \theta)] \\
H &= \nabla_x^2 \, \mathbb{E}_{q(z)}[\log p(x, z, y \mid \theta)] \bigg|_{x = x^\star}.
\end{align}
Importantly, the Hessian is block-tridiagonal such that linear solves with the Hessian and sampling from $q(x)$ scale linearly in the length of the time-series. We note that this step is not guaranteed to increase the ELBO.

Finally, the model parameters are updated by optimizing the ELBO with respect to the parameters. This corresponds to setting $\theta$ to 
\be
\theta^\star = \argmax_\theta  \mathbb{E}_{q(z) q(x)}[\log p(x, z, y \mid \theta)]. 
\ee 
We approximate this update using a single sample from $q(x)$ and marginalizing the discrete states. That is, we find $\theta^\star = \argmax_\theta  \mathbb{E}_{q(z)}[\log p(\hat{x}, z, y \,|\, \theta)]$ where $\hat{x} \sim q(x)$. The parameters are then set to a convex combination of the previous and new parameters
\begin{equation}
    \theta^{(i+1)} = \alpha \, \theta^{(i)} + (1 - \alpha) \, \theta^\star.
\end{equation}
We use either $\alpha=0.5$ or $\alpha=0.75$ in our experiments. We note that an alternative approach is to take stochastic gradient steps given samples from $q(x)$. 

This method is a generalization of the Laplace-EM method for inference in single-state linear dynamical systems with non-conjugate observations \citep{paninski2010new,macke2011empirical,macke2015estimating}. When there is only one discrete state the presented method is equivalent to Laplace-EM. We note that variational and Laplace approximations have been used together previously in other settings \citep{wang2013variational}.

A Python implementation of the vLEM algorithm for fitting (r)SLDS models is available at \url{https://github.com/slinderman/ssm}. 

\begin{figure*}[t]
    \centering
    \includegraphics[width=1.0\textwidth]{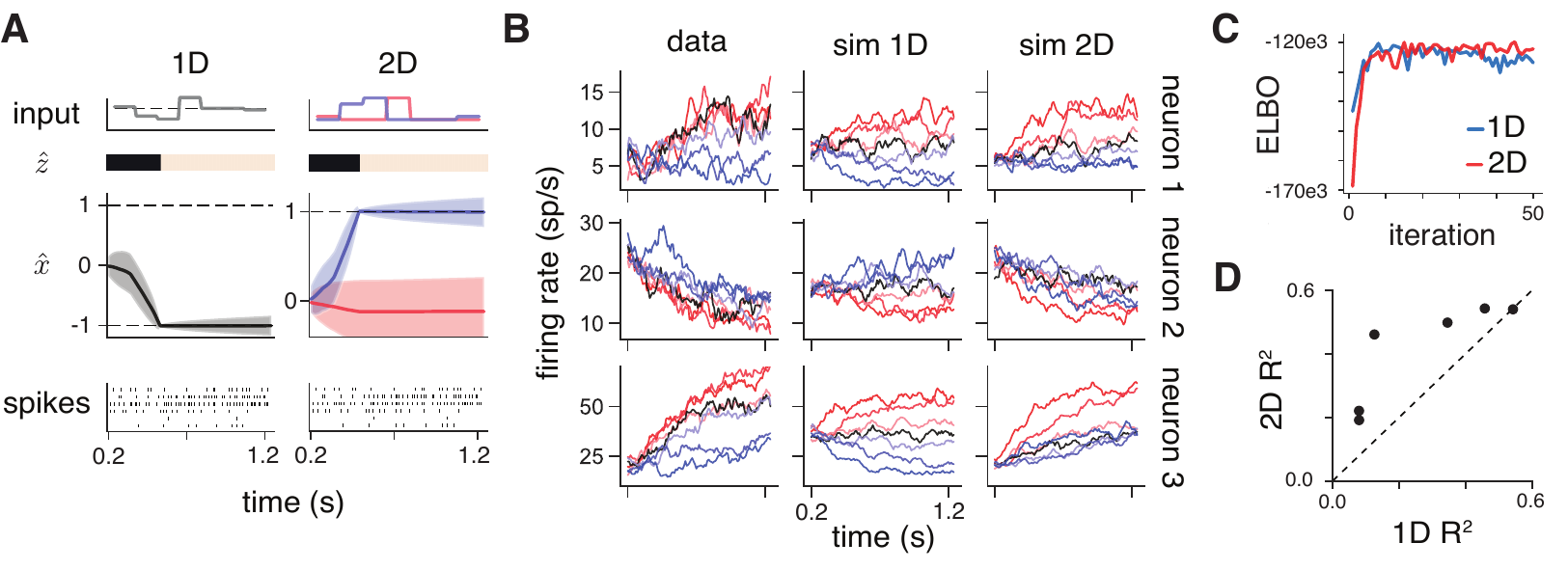}
    \caption{1D and 2D accumulation models of LIP responses during a discrete-pulse motion direction discrimination task \citep{yates2017functional}. \textbf{A.} The inferred continuous and discrete states for an example trial. \textbf{B.} The experimental and simulated trial-averaged responses across stimulus conditions. The color indicates the net direction of motion and the darkness indicates the strength of motion. \textbf{C.} The ELBO as a function of the inference iteration. \textbf{D.} The variance explained in the trial-averaged responses.}
    \label{fig:fig4}
\end{figure*}

\section{Experiments}

We demonstrate the constrained recurrent state-space modeling framework and vLEM inference with one simulated example and two applications to recordings of neural activity during decision-making tasks. 

\subsection{Simulated 2D accumulator}

We first tested the vLEM inference algorithm on data simulated from a 2D race accumulator model with Poisson spike count observations. We simulated $400$ time series (i.e. trials) with $T=100$ time points in each trial, which is similar to the amount of data collected in experiments \citep{yates2017functional}. The generated spike counts are from $N=10$ neurons. To generate realistic numbers of spike counts, we scaled the rates of the Poisson distribution to have a mean near $0.4$ spikes per $10$ms time bin. We simulated two-dimensional pulse inputs to the two accumulator dimensions. We fit the simulated data with a 2D race accumulator model using 150 iterations of the vLEM algorithm (Figure~\ref{fig:fig3}A-B). Notably, the inferred discrete and continuous states closely matched the true states. 

To test parameter recovery, we fit the model to $10$ different simulated datasets for each of 25, 50, 100, and 250 trials. We used the same parameters for each simulation and $N=10$ neurons with mean spike counts near $0.4$ spikes per bin. Importantly, the MSE between the true and generated parameters decreased as the number of simulated trials increased (Figure~\ref{fig:fig3}C). 

\subsection{1D vs. 2D accumulator models}

We used the modeling framework to compare 1D and 2D accumulator models of neural responses in the monkey lateral intraparietal area (area LIP) during a discrete-pulse accumulation task \citep{yates2017functional}. In this task, a series of seven motion pulses was presented to the animal over a period of $1050$ms. Each pulse had a variable strength of motion in one of two directions. The animal was trained to determine the net direction of motion across the seven pulses. 

We fit 1D accumulation-to-bound and 2D race accumulator models to the responses of $6$ LIP neurons simultaneously recorded during 327 trials of this task (Figure~\ref{fig:fig4}). We fit the models to the spike counts of these neurons binned in $10$ms bins from the period $200$ms after motion onset until $200$ms after motion offset. This is the time window in which the neurons putatively accumulate evidence. In the 1D model, the input was the net motion strength at each time point. In the 2D model, the input was separated into the two directions such that each dimension received only one direction of motion as input.  The ELBO converged within tens of iterations of the inference algorithm for each model (Figure~\ref{fig:fig4}C). The fits from the 1D and 2D models to an example trial are shown in Figure~\ref{fig:fig4}A. On this trial, the inferred switch between the accumulation and boundary state occurred in the same direction and at about the same time across the two models. 

The trial-averaged responses of three of the neurons are shown in Figure~\ref{fig:fig4}B, where the responses are averaged over trials with similar net motion strengths. To check how well the parameters of the inferred model correspond to the data, we simulated data from the fit models. For each stimulus in the training data we simulated spikes counts on three independent trials. We then computed the trial-averaged simulated responses in the same manner as the true responses (Figure~\ref{fig:fig4}B). The fit 1D accumulator model was unable to capture uniform increases or decreases in the trial-averaged responses across task conditions. However, the simulated firing rates from the 2D accumulator model produced responses that either decreased (neuron 2) or increased (neuron 3) across all conditions, as seen in the data. Accordingly, the 2D accumulator model explained a higher fraction of the variance in the trial-averaged responses for five of the six neurons (Figure~\ref{fig:fig4}D). Finally, we decoded choices from the inferred continuous latent variables at the final time point $T$ on each trial using the sign of $\hat{x}_T$ for the 1D accumulator model and the dimension with a larger $\hat{x}_T$ for the 2D accumulator model. In this case, the 1D model provided higher decoding accuracy (96.1\% for 1D, 91.7\% for 2D). 

\begin{figure}
    \centering
        \includegraphics[width=0.45\textwidth]{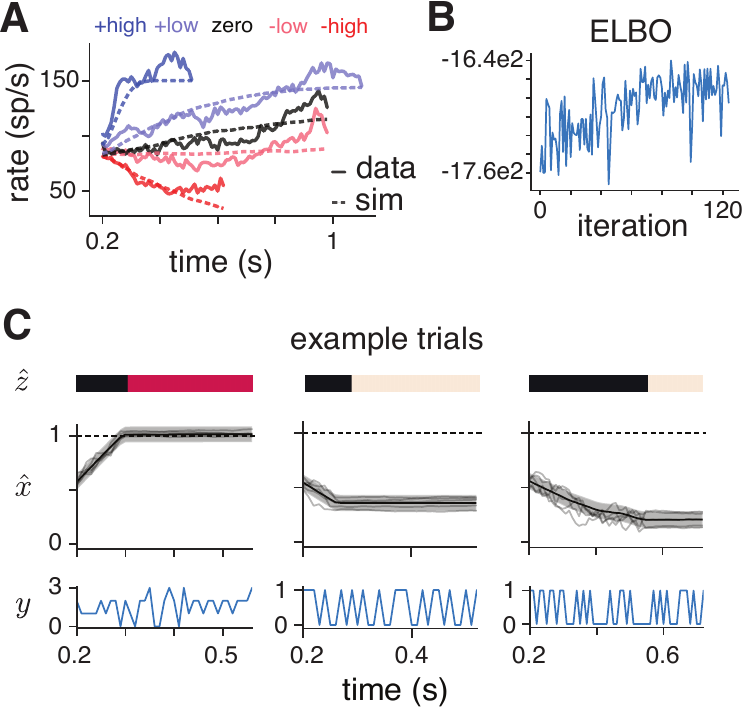}
    \caption{Ramping model with variable lower boundary fit to the responses of an LIP neuron during a motion direction discrimination task \citep{roitman2002response}. \textbf{A.} The experimental and simulated trial-averaged responses for each task condition. \textbf{B.} The ELBO as a function of inference iteration. \textbf{C.} The inferred discrete and continuous latent states on three example trials.}
    \label{fig:fig5}
\end{figure}

\subsection{Ramping model with variable lower boundary}
Since the transitions are probabilistic, we can fit a model with a variable lower boundary where switches from the accumulation state to the boundary state may occur at different values of $x$ across trials. We used this approach to fit a ramping model with a variable lower boundary to the spike count responses of an LIP neuron during a reaction-time version of the random dot motion task \citep{roitman2002response}. To do this, we introduced parameters $B_\mathsf{lb}$ and $\gamma_\mathsf{lb}$ that governed the location and sharpness of the lower boundary, and we learned these parameters during inference. We kept a sharp upper boundary fixed at $x = 1$. Example simulated trajectories from this model are shown in Figure~\ref{fig:fig2}.

We fit the model to the spiking responses of a single LIP neuron during 225 trials of a reaction time version of the random dot motion task \citep{roitman2002response}. We binned the spiking responses in $10$ms time bins and included the period $200$ms after motion onset until $50$ms before the saccade in our analysis. The trials were separated into five categories of stimulus motion. We fit the model using $125$ iterations of the vLEM algorithm. True and simulated trial-averaged responses from the fit model are shown in Figure~\ref{fig:fig5}A. We found that a variable lower boundary stopped downward sloping trajectories, as we inferred switches to the lower boundary state at different values of $x$ (Figure~\ref{fig:fig5}C). 

\section{Discussion}

We introduced a unifying framework for decision-making models based on constrained recurrent, switching linear dynamical systems. We also presented a variational Laplace EM algorithm for inference in rSLDS models. While we applied the algorithm to fit rSLDS models constrained by decision-making theory, we also expect the algorithm to be broadly useful for fitting unconstrained rSLDS models. Finally, we demonstrated our framework by fitting 1D and 2D accumulator models and a ramping model with a variable lower boundary to neural responses during decision-making. 

Our approach is an example of constraining statistical models of neural data with computational theory \citep{linderman2017using}. This is in contrast to a long line of work developing general purpose models of neural dynamics \citep{smith2003estimating,byron2009gaussian,macke2011empirical,linderman2017bayesian,wu2017gaussian,zhao2017variational,pandarinath2018inferring,duncker19interpretable}. We find the theory-driven approach to be natural for this setting, as it provides statistical tests of how well specific decision-making dynamics describe neural activity. Additionally, in our framework we can start with a theory-constrained model and add flexibility as desired or needed, for example by including additional latent dimensions or other discrete states. We look forward to exploiting this flexibility in future work. 

\subsubsection*{Acknowledgements}
We thank Jacob Yates and Alex Huk for sharing their LIP data and Jamie Roitman and Michael Shadlen for making their LIP data publicly available. We also thank Orren Karniol-Tambour for helpful discussions. D.M.Z. was supported by NIH grant T32MH065214. J.W.P. was supported by grants from the Simons Collaboration on the Global Brain (SCGB AWD543027), the NIH BRAIN initiative (NS104899 and R01EB026946), and a U19 NIH-NINDS BRAIN Initiative Award (5U19NS104648).  S.W.L. was supported by NIH grants U19NS113201 and R01NS113119.

\interlinepenalty=10000
\bibliography{dzbib.bib}


\appendix
\onecolumn
\renewcommand\thefigure{\thesection\arabic{figure}}    
\setcounter{figure}{0}  

\section{Additional Simulated Experiments}
\label{sec:appendix_experiments}

Here we demonstrate fitting models with collapsing boundaries and trial-history effects to simulated data. We also compare the variational Laplace EM inference algorithm with a black box variational inference approach. 

\subsection{Collapsing boundaries}

\begin{figure}[h!]
    \centering
    \includegraphics[width=0.95\textwidth]{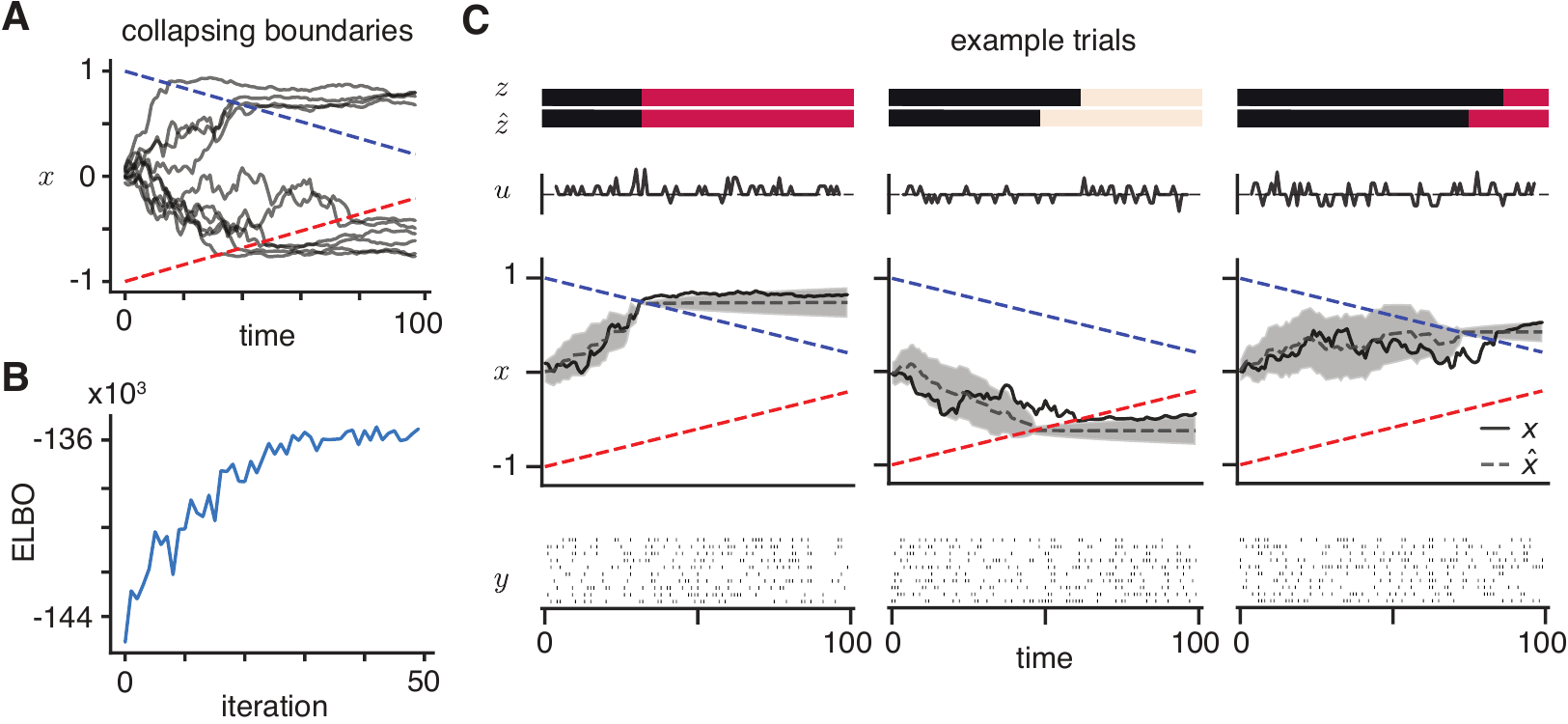}
    \caption{Simulated experiment with collapsing boundaries. \textbf{A.} Simulated latent trajectories from a one-dimensional accumulation model with collapsing upper and lower boundaries. \textbf{B.} ELBO as a function of iteration. \textbf{C.} The simulated data and the true and inferred discrete and continuous states for three example trials.}
    \label{fig:fig_collapsing}
\end{figure}

We simulated spike counts from $10$ Poisson neurons from a one-dimensional accumulation model with collapsing boundaries (Figure~\ref{fig:fig_collapsing}A). The bin size was $\Delta_t = 0.01$, the trial length was $T = 100$, and the number of trials was $200$. The inputs were the difference of two dimensional pulses. The accumulation state parameters were $V_{{\mathsf{acc}}} = 0.01$ and $\sigma_{{\mathsf{acc}}}^2 = 0.001$.

The boundaries started at $\pm 1$ and collapsed towards zero at a rate of $0.008$ per time bin, which means that the final boundary values at $T=100$ were $\pm 0.2$. We implemented this model with the following steps. First, we augmented the input vector with the current time of the trial such that $u_t = [s_t, t]^\top$ where $s_t$ is the current stimulus input. Importantly, we set the second dimension of the input weight parameter $V_{{\mathsf{acc}}}^{(2)}$ to zero so the time is not input to the continuous dynamics $x$. We modified the transitions to depend on the input with the following parameterization
\be
p(z_{t} \mid z_{t-1}, x_{t-1}) \propto \exp \left \{ \gamma (R_{z_{t-1}} + r \, x_{t-1} + W u_t) \right \}, \quad W = \begin{bmatrix}
0 & 0 \\
0 & \beta \\
0 & \beta
\end{bmatrix}
\ee
where $\beta$ is a scalar parameter that controls the slope of the boundary. We set the left column of $W$ to zeros so the sensory input does not directly affect the transitions. We set $\beta = 0.008$, which corresponds to the rate of the collapsing boundaries as described above. We note that asymmetric collapsing boundaries can be implemented by having separate $\beta$ parameters for each dimension. While we fix the slope parameter $\beta$, its value could be learned. The parameters $R_{z_{t-1}}$ and $r$ have the same form as in the original 1D accumulator model.

We fit the collapsing boundaries model to the simulated data using 50 iterations of the vLEM algorithm (Figure~\ref{fig:fig_collapsing}B). The inferred continuous and discrete states from the algorithm were similar to the true latent states (Figure~\ref{fig:fig_collapsing}C).

\subsection{Trial-history}

\begin{figure}[h!]
    \centering
    \includegraphics[width=0.95\textwidth]{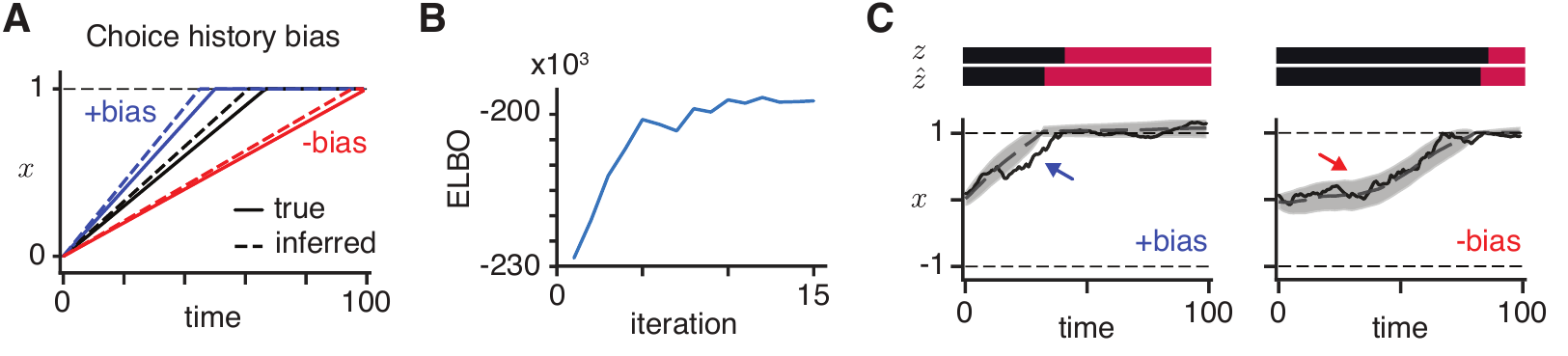}
    \caption{Simulated experiment with trial-history effects. \textbf{A.} True and inferred averaged drift rates for positive going trials with and without the choice history bias. The previous choice biases the drift rate upwards (previous choice corresponds to upper boundary) and downwards (previous choice corresponds to lower boundary). \textbf{B.} The ELBO as a function of optimization iteration. \textbf{C.} True and inferred states for positive going trials with positive (\textit{left}) and negative (\textit{right}) biases.}
    \label{fig:fig_history}
\end{figure}

The modeling framework allows for trial-history effects based on the previous choice, reward, or stimulus. Here we simulated data from a model where the previous trial choice affects the drift rate (Figure~\ref{fig:fig_history}). We implemented this by including the previous trial choice $c_\mathsf{prev} = \{-1, 1\}$ as an additional input covariate. The input at each time point on a given trial was $u_t = [s_t, c_\mathsf{prev}]$. In this case, we learn each dimension of the input weights $V_{{\mathsf{acc}}} \in \mathbb{R}^2$. The element in the second dimension corresponds to the bias in the drift rate. This parameterization enforces a symmetric drift bias, but it is again possible to relax the symmetry. 

We simulated spike counts of 5 Poisson neurons from this model with a bin size $\Delta = 0.1$. Each trial had length $T = 100$ and we simulated $N = 200$ trials. In this simulation, the input on each trial was a constant drift of $s_t = 0.015$ for positive going trials and $s_t = -0.015$ for negative going trials. The drift bias was $0.005$ and the variance was $\sigma_{{\mathsf{acc}}}^2 = 0.001$. The average drift rate on positive going trials is shown in Figure~\ref{fig:fig_history}. The bias increased the average drift when the previous choice was $+1$ (blue line) and decreased the average drift when the previous choice was $-1$. 

We fit this model using 15 iterations of the vLEM algorithm (Figure~\ref{fig:fig_history}B). The inferred drift rates in the fit model were similar to the true drift rates (Figure~\ref{fig:fig_history}A). Additionally, the inferred latent states correctly followed the bias shown in the true latent states (Figure~\ref{fig:fig_history}C).

\subsection{Comparison of vLEM and BBVI}

\begin{figure}[h!]
    \centering
        \includegraphics[width=0.95\textwidth]{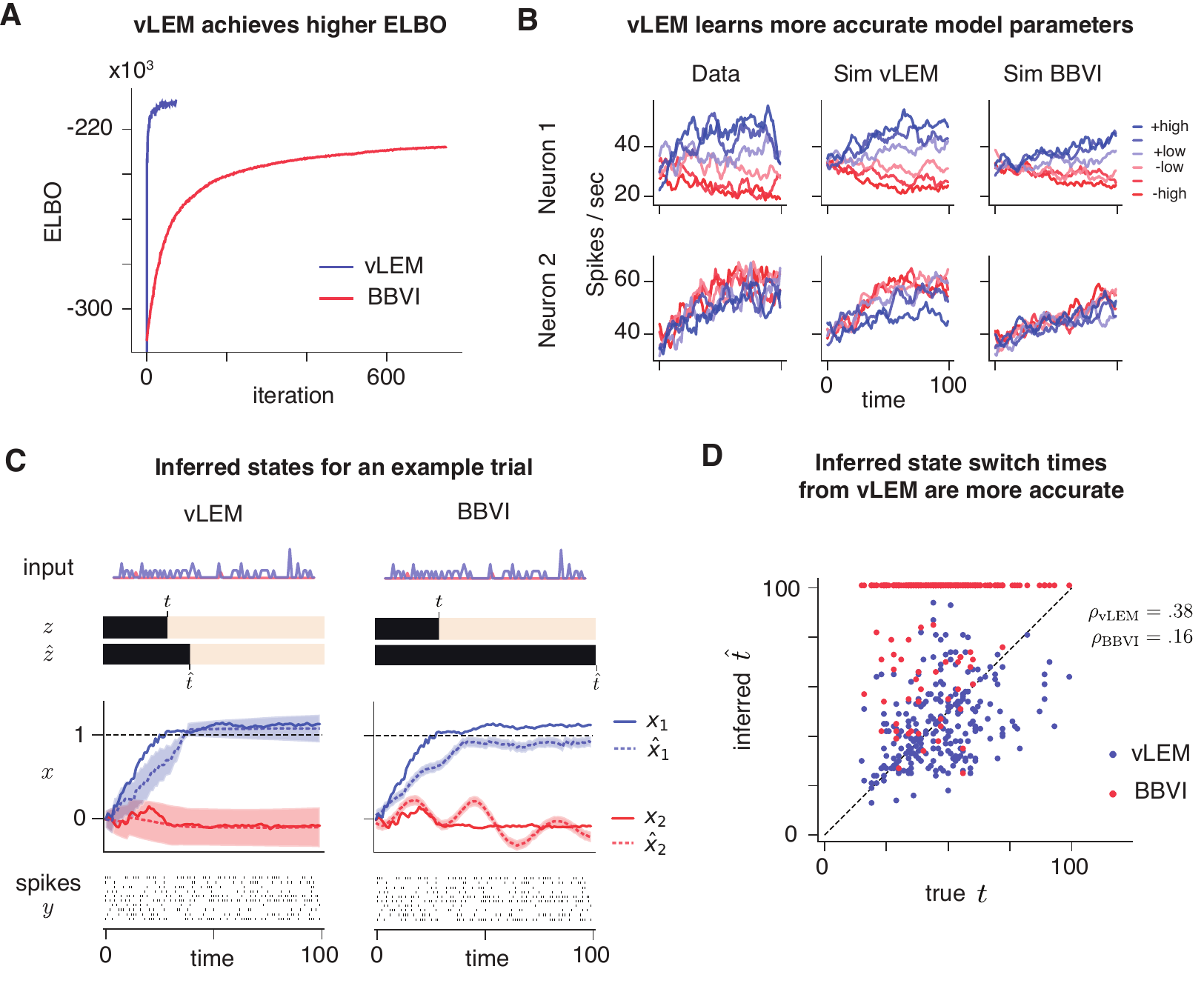}
    \caption{Comparison of vLEM and BBVI for fitting a 2D accumulator model with Poisson neurons. \textbf{A}. The ELBO as a function of algorithm iteration. \textbf{B.} The true average neural responses across different evidence strengths (line colors) and simulated responses from the fit model using vLEM or BBVI. Here, “+high” is strong stimulus motion towards the “+” direction while “-high” is strong stimulus motion to the opposite “-” direction. \textbf{C.}  The inferred (dashed lines, $\hat{z}$, and $\hat{x}$) and true (solid lines,  $z$, and $x$) continuous and discrete states using vLEM and BBVI for one typical example trial. The true $(t)$ and inferred $(\hat{t})$ transition times are noted on the colored bars. \textbf{D.} The true and inferred transition times from the accumulation state to the boundary state for vLEM and BBVI for all trials.}
    \label{fig:comp_bbvi_lds}
\end{figure}

To demonstrate the utility of vLEM for our models, we compared using vLEM and black box variational inference (BBVI) to fit a 2D race accumulator model. We simulated 10 Poisson neurons across 250 trials from the 2D race accumulator model with a bin size of $\Delta_t = 0.01$. We note this is the same model as that in Figure~\ref{fig:fig3}. 

We fit a 2D race accumulator model to the simulated data using vLEM and BBVI. For BBVI, we used a jointly Gaussian posterior over the continuous latent variables with block-tridiagonal structure in the precision of the covariance matrix and we marginalized the discrete states  \citep{archer2015black,gao2016linear,linderman2019hierarchical}. We initialized the models with the same parameters and with the same posterior over the continuous latent variables. 

The results of fitting the model with vLEM and BBVI are shown in Figure~\ref{fig:comp_bbvi_lds}. First, we found that vLEM achieved substantially higher ELBO values (Figure~\ref{fig:comp_bbvi_lds}A). Next, the learned model parameters from vLEM generated data that are more similar to the true simulated data than BBVI (Figure~\ref{fig:comp_bbvi_lds}B). This is shown by the similarity in the PSTHs in the first two columns. 

Crucially, vLEM provided more accurate inferences about the latent states (Figure~\ref{fig:comp_bbvi_lds}C). BBVI had difficulty learning transitions from accumulation to boundary and had qualitatively poorer uncertainty estimates (Figure~\ref{fig:comp_bbvi_lds}C-D). On many trials BBVI did not infer a switch from accumulation to boundary. 

\clearpage
\newpage

\section{Variational Laplace-EM Inference}
\label{sec:appendix_inference}

Here we describe in more detail the variational Laplace-EM inference method. As noted in the main text, we introduce a factorized approximate posterior $q(z) q(x) \approx p(z,x \mid y, \theta)$ over the discrete and continuous latent variables. With those distributions we lower-bound the marginal likelihood with
\begin{align*}
\mathcal{L}_q(\theta) & = \mathbb{E}_{q(z) q(x)}[\log p(x, z, y \mid \theta) - \log q(z) q(x)] \\
& = \mathbb{E}_{q(z) q(x)}[\log p(x, z, y \mid \theta)] - \mathbb{E}_{q(z)}[\log q(z)] - \mathbb{E}_{q(x)}[\log q(x)].
\end{align*}
To optimize this objective, we alternate between updating 1) $q(z)$, 2) $q(x)$ and 3) $\theta$. The updates to $q(z)$ and $\theta$ follow from optimizing the lower bound $\mathcal{L}_q (\theta)$. The update to $q(x)$ is an approximate update and is therefore not guaranteed to increase the value of the lower bound. 

\subsection{Update discrete state posterior}
We update $q(z)$ via the optimal coordinate ascent variational inference update 
\be
q^\star(z)  \propto \exp ( \mathbb{E}_{q(x)}[\log p(x, z, y \mid \theta)] ).
\ee
To compute this, we expand the expected log joint probability
\begin{align*}
\mathbb{E}_{q(x)}[\log p(x, z, y \mid \theta)]  & = \mathbb{E}_{q(x)} \bigg[ \log p(z_1, x_1 \mid \theta) + \sum_{t=2}^{T} \log p(x_t \mid x_{t-1}, z_t, \theta) \\
&\qquad\qquad + \sum_{t=1}^{T-1} \log p(z_{t+1} \mid z_t, x_t, \theta) + \sum_{t=1}^T \log p(y_t \mid x_t, z_t, \theta) \bigg] \\
& = \phi(z_1, x_1) + \sum_{t=2}^{T} \phi(z_t, x_t, x_{t-1}) + \sum_{t=1}^{T-1} \phi(z_t, z_{t+1}, x_t) + \sum_{t=1}^T \phi(z_t, x_t, y_t)
\end{align*}
where we have introduced the potentials 
\begin{align*}
\phi(z_1, x_1) & =  \mathbb{E}_{q(x)}[\log p(z_1, x_1 \mid \theta)] \\
\phi(z_t, x_t, x_{t-1}) & = \mathbb{E}_{q(x)} [\log p(x_t \mid x_{t-1}, z_t, \theta)] \\
\phi(z_t, z_{t+1}, x_t) & =  \mathbb{E}_{q(x)} [\log p(z_{t+1} \mid z_t, x_t, \theta) ] \\
\phi(z_t, x_t, y_t) & = \mathbb{E}_{q(x)} [\log p(y_t \mid x_t, z_t, \theta)].
\end{align*}
We used samples from $q(x)$ to estimate the expectations in these potentials. We used a default of a single sample in our simulations and applications to data. We note that if the observations are independent of the discrete states when conditioned on the continuous states (i.e.~$\log p(y_t \mid x_t, z_t, \theta) = \log p(y_t \mid x_t, \theta)$) 
then the emission potential $\phi(z_t, x_t, y_t)$ can be disregarded for updating $q(z)$.

We introduce the normalizing constant $Z(\theta)$ of the distribution such that
\be
q(z) = \frac{1}{Z(\theta)} \exp \bigg(  \phi(z_1, x_1) + \sum_{t=2}^{T} \phi(z_t, x_t, x_{t-1}) + \sum_{t=1}^{T-1} \phi(z_t, z_{t+1}, x_t) + \sum_{t=1}^T \phi(z_t, x_t, y_t) \bigg).
\ee
Conditioned on the estimates of the potentials, we have a factor graph equivalent to the factor graph of an HMM. Therefore we compute the unary and pairwise marginals over $z$ and the normalizing constant using the forward-backwards algorithm. 
We evaluate the entropy term in the ELBO using the potentials, the unary and pairwise marginals, and the normalizing constant as
\begin{align*}
\mathbb{E}_{q(z)}[\log q(z)] & = \mathbb{E}_{q(z)}\bigg[- \log Z(\theta) + \phi(z_1, x_1) + \sum_{t=2}^{T} \phi(z_t, x_t, x_{t-1}) + \sum_{t=1}^{T-1} \phi(z_t, z_{t+1}, x_t) + \sum_{t=1}^T \phi(z_t, x_t, y_t) \bigg] \\
& = -\log Z(\theta) + \mathbb{E}_{q(z)} [\phi(z_1, x_1)] + \sum_{t=2}^{T} \mathbb{E}_{q(z)}  [ \phi(z_t, x_t, x_{t-1}) ] + \sum_{t=1}^{T-1} \mathbb{E}_{q(z)} [\phi(z_t, z_{t+1}, x_t)] \\
& \quad + \sum_{t=1}^T  \mathbb{E}_{q(z)} [\phi(z_t, x_t, y_t)].
\end{align*}

\subsection{Update continuous state posterior}
We update $q(x)$ with a Laplace approximation around the mode of $\mathbb{E}_{q(z)}[\log p(x, z, y \mid \theta)]$ such that
\begin{align*}
q^\star(x)  &= \mathcal{N}(x^\star, -H^{-1}) \\
x^\star &= \argmax_x \mathbb{E}_{q(z)}[\log p(x, z, y \mid \theta)] \\
H &= \nabla_x^2 \mathbb{E}_{q(z)}[\log p(x, z, y \mid \theta)] \bigg|_{x = x^\star}.
\end{align*}
To compute the Hessian we expand the terms in the objective
\begin{align*}
\mathcal{L}(x) &= \mathbb{E}_{q(z)}[\log p(x, z, y | \theta)] \\
& = \mathbb{E}_{q(z)} \bigg[ \log p(z_1 \mid \theta) +  \log p(x_1 \mid z_1, \theta) + \sum_{t=2}^{T} \log p(x_t \mid x_{t-1}, z_t, \theta) \\
&\qquad \qquad + \sum_{t=1}^{T-1} \log p(z_{t+1} \mid z_t, x_t, \theta) + \sum_{t=1}^T \log p(y_t \mid x_t, z_t, \theta) \bigg] \\
& = \phi(x_1, z_1) + \sum_{t=2}^T \phi(x_t, x_{t-1}, z_t) + \sum_{t=1}^{T-1} \phi(x_t, z_t, z_{t+1}) + \sum_{t=1}^T \phi(x_t, y_t, z_t) + \text{const} 
\end{align*}
where
\begin{align*}
\phi(x_1, z_1) & =  \mathbb{E}_{q(z)}[\log p(x_1 \mid  z_1 , \theta)] = \sum_k q(z_1 = k) \log p(x_1 \mid z_1 = k, \theta) \\
\phi(x_t, x_{t-1}, z_t) & = \mathbb{E}_{q(z)}[ \log p(x_t \mid x_{t-1}, z_t, \theta) ] = \sum_{k} q(z_t = k) \log p(x_t \mid x_{t-1}, z_t = k, \theta)  \\
\phi(x_t, z_t, z_{t+1}) & = \mathbb{E}_{q(z)}[  \log p(z_{t+1} \mid z_t, x_t, \theta) ] = \sum_k \sum_j q(z_t = k, z_{t+1} = j) \log p(z_{t+1} = j \mid z_t = k, x_t, \theta) \\  
\phi(x_t, y_t, z_t) & = \mathbb{E}_{q(z)}[  \log p(y_t \mid x_t, z_t, \theta) ] = \sum_k q(z_t = k) \log p(y_t \mid x_t , z_t = k, \theta) .
\end{align*}
The above derivation was written in full generality. If the emission potential does not depend on the discrete state then the emission potential simplifies to~$\phi(x_t, y_t, z_t) =  \log p(y_t \mid x_t, \theta)$. 
Also, if there are no recurrent dependencies (as in a standard SLDS) then the transition term  $\log p(z_{t+1} \mid z_t, x_t, \theta)$ is equal to $\log p(z_{t+1} \mid z_t, \theta)$ and therefore the transition potential $\phi(x_t, z_t, z_{t+1})$ no longer depends on $x_t$. 

We require the Hessian matrix for the Laplace approximation. This matrix is given by
\begin{align*}
\nabla_x^2 \mathcal{L}(x) & = \nabla_x^2 \mathbb{E}_{q(z)}[\log p(x, z, y \mid \theta)] \\
& = \nabla_x^2 \phi(x_1, z_1) + \sum_{t=2}^T \nabla_x^2 \phi(x_t, x_{t-1}, z_t) + \sum_{t=1}^{T-1} \nabla_x^2 \phi(x_t, z_t, z_{t+1}) + \sum_{t=1}^T \nabla_x^2 \phi(x_t, y_t, z_t) 
\end{align*}
where
\begin{align*}
\nabla_x^2 \phi(x_1, z_1) & = \sum_k q(z_1 = k) \nabla_x^2 \log p(x_1 \mid z_1 = k, \theta) \\
\nabla_x^2 \phi(x_t, x_{t-1}, z_t)  & = \sum_{k} q(z_t = k) \nabla_x^2  \log p(x_t \mid x_{t-1}, z_t = k, \theta)  \\
\nabla_x^2 \phi(x_t, z_t, z_{t+1}) & = \sum_k \sum_j q(z_t = k, z_{t+1} = j) \nabla_x^2 \log p(z_{t+1} = j \mid z_t = k, x_t, \theta) \\
\nabla_x^2 \phi(x_t, y_t, z_t) & = \sum_k q(z_t = k) \nabla_x^2 \log p(y_t \mid x_t , z_t = k, \theta).
\end{align*}
Therefore, we can compute the Hessian by computing the contributions to the Hessian of the dynamics, emission, and transition potentials. 

The Hessian has size $TD \times TD$ for a time series of length $T$ with latent dimensionality $D$ but has a sparse, block tridiagonal structure with blocks of size $D \times D$. The terms in the Hessian from the initial state, transition, and emission potentials only contribute terms to the primary block diagonal. The dynamics potentials contribute terms to both the primary and first off-diagonal blocks. Throughout, we only represent and store the main and lower diagonal blocks of the Hessian. This reduces storage from the full $(TD)^2$ to $(2T-1)D^2$ such that it is linear in $T$. For linear solves and matrix inversions of the Hessian, we also use algorithms that exploit the block tridiagonal structure. 

To find the most likely latent path $x^\star$, we use Newton's method with a backtracking line search. However, we can also use optimization routines that require only gradient information (lBFGS) or require only gradient information and Hessian-vector products (Newton-CG or trust-region Newton-CG). 


\subsection{Update parameters}
We update the model parameters by approximately optimizing the ELBO with respect to the parameters
\be
\theta^\star = \argmax_\theta  \mathbb{E}_{q(z) q(x)}[\log p(x, z, y \mid \theta) - \log q(z) q(x)]. 
\ee 
Instead of optimizing the expectation under the full distribution of $q(x)$ we optimize
\be
\theta^\star = \argmax_\theta \mathbb{E}_{q(z)}[\log p(\hat{x}, z, y \mid \theta)]
\ee
where $\hat{x}$ is a sample from $q(x)$ and we have dropped terms that do not depend on $\theta$. Conditioned on $\hat{x}$, the update consists of M-steps on the transition, dynamics, and emission parameters. We use either exact updates (where applicable) or lFBGS to implement the M-steps. Finally, we set the parameters at iteration $i$ via a convex combination of the new parameters $\theta^\star$ and the parameters at the previous iteration
\be
\theta_i = (1 - \alpha) \, \theta^\star + \alpha \, \theta_{i-1}.
\ee
We note that we can also update the parameters using stochastic gradient ascent with samples from $q(x)$.

\subsection{Initialization}

We can exploit the known structure of the 1D and 2D accumulation-to-bound models to initialize some of the parameters. For the 1D and 2D accumulation-to-bound models we set the emission parameter $d$ to the mean spike counts across trials in the first three time bins. In the 1D model, we set the emission parameter $C$ using the firing rate at the end of trials with strong input to the upper ($
\lambda_\mathsf{UB}$) and lower ($\lambda_\mathsf{LB}$) boundaries. Given those values for each neuron and the fact that the boundaries are at $\pm 1$ for this model, we set $C = \frac{1}{2} (\lambda_\mathsf{UB} - \lambda_\mathsf{LB})$. In the 2D model, for each neuron we initialized the elements of $C$ as the difference between the firing rate at the end of trials with strong net input and the mean rate $d$. We did this for each dimension of the input and corresponding element in $C$. For the models and data in this paper, we did not identify procedures to reliably estimate the initial underlying latent dynamics parameters. Therefore we randomly initialized the input weights and dynamics variance and set the initial dynamics matrix to $A_\mathsf{acc} = I$.  




\end{document}